\documentclass[aps,prd,nofootinbib]{revtex4-2}
\usepackage[colorlinks=true, pdfstartview=FitV, linkcolor=blue, citecolor=red, urlcolor=magenta]{hyperref}
\usepackage{graphicx}
\usepackage{latexsym}
\usepackage{amsmath}
\usepackage{amsfonts}
\usepackage{amssymb}
\usepackage{verbatim}
\usepackage[utf8]{inputenc}
\usepackage[T1]{fontenc}

\newcommand{\be}{\begin{equation}}
\newcommand{\ee}{\end{equation}}
\newcommand{\bea}{\begin{eqnarray}}
\newcommand{\eea}{\end{eqnarray}}

\newcommand{\ben}{\begin{eqnarray}}
\newcommand{\een}{\end{eqnarray}}

\begin{document}

\title{Aether-Electromagnetic theory and the Casimir effect}


\author{$^{1,2}$K. E. L. de Farias}
\email{klecio.lima@uaf.ufcg.edu.br}

\author{$^{1}$M. A. Anacleto}
\email{anacleto@df.ufcg.edu.br}

\author{$^{1}$E. Passos}
\email{passos@df.ufcg.edu.br}

\author{ $^{3}$Iver Brevik}
\email{iver.h.brevik@ntnu.no}

\author{$^{2}$ Herondy Mota}
\email{hmota@fisica.ufpb.br}

\author{$^{1}$Jo\~ao R. L. Santos}
\email{joaorafael@df.ufcg.edu.br}


\affiliation{$^{1}$Departamento de F\'{\i}sica, Universidade Federal de Campina Grande,\\
Caixa Postal 10071, 58429-900, Campina Grande, Para\'{\i}ba, Brazil.}
\affiliation{$^{2}$Departamento de F\' isica, Universidade Federal da Para\' iba,\\  Caixa Postal 5008, Jo\~ ao Pessoa, Para\' iba, Brazil.}
\affiliation{$^{3}$Department of Energy and Process Engineering, Norwegian University of Science and Technology\\  N-7491 Trondheim, Norway.}


\begin{abstract}
In this study, we explore the impact of an additional dimension, as proposed in Kaluza-Klein's theory, on the Casimir effect within the context of Lorentz invariance violation (LIV), which is represented by the ``aether field''. We demonstrate that the Casimir energy is directly influenced by the presence of the fifth dimension, as well as by the aether parameter. Consequently, the force between the plates is also subject to variations of these parameters. Furthermore, we examine constraints on both the size of the extra dimension and the aether field parameter based on experimental data. The LIV parameter can provide insights into addressing the size-related challenges in Kaluza-Klein's theory and offers a mean to establish an upper limit on the size of the extra dimension. This helps to rationalize the difficulties associated with its detection in current experiments.
\end{abstract}
\pacs{11.15.-q, 11.10.Kk}
\maketitle


\section{Introduction}

The possibility of extra dimensions has gained prominence in current physics models. It is often investigated as a solution to open problems that the standard model cannot answer. These problems include quantum gravity since attempts to quantize gravity imply the existence of extra dimensions as a condition, the presence of dark matter in the universe, which remains poorly understood and undetected, and dark energy, which permeates the majority of the Universe. In this sense, Theodor Kaluza in 1921 \cite{Kaluza:1921tu} and Oskar Klein in 1926 \cite{Klein:1926tv}, propose a theory (KK theory) that posits the existence of an extra dimension - a fifth dimension - in an attempt to find a unified theory for the fundamental forces of nature, specifically electromagnetism and gravity. This extra dimension is compactified, small enough to be hidden at macroscopic scales.

Despite the recent attention, there is a problem involving the length, denoted here by $b$, of the extra dimension. In many theories, including the KK theory, the length $b$ is predicted to be so small that it would make detection impossible, as it suggests a length near $10^{-35}m$ (the Planck scale). At this length, only particles with energy near the Planck energy scale can access the extra dimension. To address this problem, Sean Carroll and Heywood Tam, propose the aether compactification theory as an alternative way to compactification in the presence of large extra dimensions without the introduction of branes to control the influence of a five-dimensional bulk space-time on the field content in our four-dimensional Universe. Differently from braneworlds, in this type of compactification, there are no corrections to the four-dimensional Newtonian law \cite{Carroll:2008pk}. However, the possibility of suppressing the Kaluza-Klein modes makes the mechanism efficient to guarantee a four-dimensional effective theory describing the physics of our Universe after the compactification process.

In contrast to the old aether theories that proposed the existence of a medium or substance governing the propagation of electromagnetic waves or gravitational force, the present aether approach consists of a five-dimensional Lorentz-violation tensor field that interacts only with the extra compactified dimension. This field is usually called the ``aether field'' in the literature \cite{Carroll:2008pk,Chatrabhuti:2009ew,Latta:2016jix,Santos:2017yov,Dimakis:2020dqs,Oliveira:2020chr}, and this term will be used throughout the work. Furthermore, the interaction caused by the presence of the aether modifies the dispersion relation of the other fields, as shown in Ref. \cite{Carroll:2008pk}.

The Lorentz-violating aether fields along extra dimensions affect the conventional Kaluza-Klein compactification scheme since their interactions play a fundamental role in the mass splitting of the KK towers: the mass spacings between different states are now modified. That is, the large extra dimensions, even for very high KK modes could, in principle, be accessible to a four-dimensional observer. Several studies with the aether field, both in four dimensions and in extra dimensions, have been developed
over the past few years \cite{Obousy:2008xi,Obousy:2008ca,Chatrabhuti:2009ew,
Furtado:2016zqi,Santos:2017yov,CcapaTtira:2010ez,Carroll:2009en,ADantas:2023wxd}. The main goal of the present study is to investigate the implications, of the Casimir effect, of the aether field with expectation value aligned along the extra dimension.

One of the models frequently employed in the literature, and among the simplest, involves an extra dimension compactified into a circle of length $b$ \cite{Greene:2007xu,Feng:2013zza,Erdas:2021fwe}. The mere presence of this single extra dimension provides us with ways to detect it indirectly. One phenomenological manifestation of this extra dimension can be observed in the vacuum energy density, which can be modified in order to be appreciable at a potentially detectable scale. The method to detect this energy is through the famous Casimir effect \cite{Casimir:1948dh}, where it exerts an attractive or repulsive force between plates due to the finite difference in the vacuum energy density found through regularization and renormalization procedures which subtract the infinite energy of the vacuum oscillations. As it can be measured, the influence of the extra dimension may either increase or decrease this force since the Casimir effect is a dimension-dependent phenomenon.
To observe this effect, we place parallel plates in close proximity, with a maximum distance on the order of micrometres ($\mu$m), such that the gauge field satisfies a boundary condition in the direction of the plates. Any boundary condition can, in principle, alter the vacuum frequencies of a quantum field and help understand how a system is influenced by it through the Casimir energy. Therefore, when an additional boundary condition is imposed in a different dimension of the system, as it is in our case, the resulting contribution will also influence the quantum vacuum fluctuations and, as a consequence, the Casimir (vacuum) energy \cite{lukosz1971electromagnetic,ambjorn1983properties,Bordag:2009zz}.

Following this concept, considering also the existence of a compactified extra dimension, as explored in the literature \cite{Erdas:2021fwe, Poppenhaeger:2003es, pascoal2008estimate}, it will leave a signature in the form of a contribution to the Casimir energy between the parallel plates. This signature may provide insights into the existence of an extra dimension and the nature of its contribution. Many other applications involving the Casimir effect and extra dimensions have been developed, such as in Refs. \cite{Greene:2007xu, Wongjun:2013jna}, where the authors investigated the stabilization of an extra dimension through the Casimir effect using compactification in the fifth dimension, and in \cite{Obousy:2008ca, Chatrabhuti:2009ew}, where, in addition to studying stability, they also considered the aether parameter.

Assuming the existence of an extra dimension, our proposal for the Casimir effect in a Kaluza-Klein scenario considers the Neumann boundary condition applied on the plates placed in the $z$-direction. In our model, the extra dimension is compactified by imposing a quasi-periodic boundary condition, which is mathematically described by $A(t,x, y, z) = e^{-2 \pi \beta i} A(t,x, y, z, x^5+b)$. This leads us to a generalized expression dependent on the parameter $\beta$, assuming values in the interval $0 \leq \beta < 1$. This interval includes both periodic ($\beta=0$) and antiperiodic ($\beta=1/2$) boundary conditions.

Additionally, we aim to establish new constraints on the aether field parameter, which is introduced as a possible solution to the extra dimension's length. To test these constraints, we can utilize experimental data obtained by Bressi et al. \cite{Bressi:2002fr}, where they measured the residual squared frequency shift $\Delta\nu^2$ in the context of the Casimir effect between parallel surfaces. In this experiment, a silicon cantilever coated with chromium interacts with a rigid surface, experiencing shifts in frequency due to the Casimir force. The experimental range covered distances between $0.5-3.0 \mu \mathrm{m}$, providing a high level of precision.

We introduce the aether approach in Section \ref{sec2}, where we first provide a brief description of the aether dynamic scenery. Following that, we consider the incorporation of the aether gauge coupling into the Kaluza-Klein theory. We consider the introduction of two distinct boundary conditions: the Neumann boundary condition on the plates placed in the $z$-dimension and the quasi-periodic boundary condition in the extra dimension. In Section \ref{sec3}, we derive the Casimir energy in terms of the boundary condition parameters and also explore the influence of the compactified extra dimension. Section \ref{sec4} provides the closed expression for the Casimir force and analyzes the properties and phenomenology resulting from the system configuration. In Sec. \ref{sec5}, we estimate the size of the extra dimension based on experimental data and place constraints on the aether ratio parameter. Finally, in Section \ref{sec6}, we present the conclusions drawn from the results obtained in this work. For the most part of this paper, we use natural units $\hbar=c=1$.

\section{LIV Aether approach}

In this section, we provide a brief description of the ``aether field'' approach in the dynamic context and when coupled to a gauge field. Specifically, we consider a five-dimensional flat spacetime with metric signature $(-,+,+,+,+)$ and coordinates $x^a=\left(x^\mu, x^5\right)$, where the fifth dimension is compactified on a circle of radius $R$.

\subsection{LIV Aether dynamics}
\label{sec2}

Specifically, the fifth dimension is accessed through a spacelike vector $u^a$, referred to as the aether (also known in the literature as the ``aether field''), and due to this vector, the Lorentz symmetry is spontaneously broken. Moreover, we can define a ``field strength'' tensor
\begin{equation}
    V_{a b}=\nabla_a u_b-\nabla_b u_a .
    \label{ae1}
\end{equation}
It should be highlighted that there is no relation between the above field and the electromagnetic vector potential $A_a$ or the field strength $F_{a b}=\nabla_a A_b-\nabla_b A_a$. Furthermore, the dynamic of the vector $u^a$ does not respect the $U(1)$ group of gauge transformations. The following action considers a Maxwell-type kinetic term that is used to fix a constant norm for the aether field
\begin{equation}
    S=\int d^5 x \sqrt{-g}\left[-\frac{1}{4} V_{a b} V^{a b}-\lambda\left(u_a u^a-v^2\right)+\sum_i \mathcal{L}_i\right] .
    \label{ae2}
\end{equation}
Note that $\lambda$ acts as a Lagrangian multiplier, imposing constraints on the aether field
\begin{equation}
u^a u_a=v^2.
\label{ae3}
\end{equation}
There is a formal similarity between this kind of Lorentz-violating theory and the relativistic macroscopic theory of a continuous medium: in both cases, there is introduced a constant vector, here simply called  $u^\alpha$,  which plays a distinguished role. In the latter electrodynamic case, it is simply the uniform four-dimensional velocity of the medium. It is {\it timelike}, reducing to a unit vector in the rest frame of the medium. In the present Lorentz-breaking case, the vector  $u^\alpha$ is spacelike, as shown in Eq. \eqref{ae3}. Any observable effect critically dependent on this five-vector would imply serious physical consequences. It is observed that both in the four- and the five-dimensional cases, the appearance of these constant vectors can be done via Lagrangian multipliers in the action integral. Readers who might be interested in covariant electrodynamics of media may consult, for instance, \cite{brevik1970electromagnetic,brevik2020classical}.

The last term in the action, the sum $\mathcal{L}_i$ represents interaction terms coupled to the aether field. These terms are matter fields which will be commented on later. The equation of motion for the aether field, ignoring the interaction terms, is given by
\begin{equation}
\nabla_a V^{a b}+v^{-2} u^b u_c \nabla_d V^{c d}=0
\label{ae4}
\end{equation}
There is the trivial solution  $V_{ab}=0$ which solves the equation of motion for any configuration, and there is a particular background solution related to the aether field of the form
\begin{equation}
    u^a=(0,0,0,0,v),
    \label{ae5}
\end{equation}
where the fifth component of the vector is the only non-zero component to preserve Lorentz invariance in a 4-dimensional non-compact space. This configuration will be used for the rest of the work.

\subsection{Aether-gauge field coupling}

At this point, we derive the interaction between the aether and the gauge field by considering the five-dimensional electromagnetic Lagrangian with an extension of the lowest-order coupling to the aether background field $u^a$, given as: \cite{Carroll:2008pk}
\begin{equation}
    \mathcal{L}_{A}=-\frac{1}{4} F_{a b} F^{a b}+\frac{1}{2 \mu_A^2} u^a u^b \eta^{c d} F_{a c} F_{b d},
    \label{el1}
\end{equation}
where $u^a=\left(u^\nu, u^5\right)$, $\nu=0,\ldots,3$ and $\eta^{cd}$ is the Minkowski metric. The five-dimensional field strength tensor is defined in terms of the potential $A_a$ and remains with the usual relation $F_{a b}=\partial_a A_b-\partial_b A_a$. Equation \eqref{el1} stays invariant under the gauge transformation: $\delta A^a=\partial^a \phi(x)$. From Equation \eqref{el1}, we may find the equation of motion in five dimensions, given by
\begin{equation}
    \partial_a F^{a b}=\mu_A^{-2}\left(u_c u^b \partial_a F^{c a}-u_c u^a \partial_a F^{c b}\right).
    \label{el2}
\end{equation}
The above equation can be rewritten using the Lorentz gauge, $\partial_aA^a=0$ and the axial gauge, $u_aA^a=0$, Eq. \eqref{el2} is then reduced to
\begin{align}
\left(\partial_a \partial^a+\mu_A^{-2} u^c u^a \partial_{a} \partial_c\right) A^b=0.
\label{el3}
\end{align}
Going to Fourier space, $A^\nu \propto$ $ e^{i k_\mu x^\mu+i k_5 x^5}$, and consider an aether background of the form $u^a=(0,0,0,0, v)$ as justified by Eq. \eqref{ae5}, we get as result
\begin{align}
 k_\mu k^\mu=-\left(1+\alpha_A^2\right) k_5^2,
 \label{el4}
\end{align}
where $\alpha_A=v / \mu_A$ is a dimensionless parameter.

It is straightforward to note that
in the absence of the extra dimension, the dispersion relation \eqref{el4} leads us to the energy of a photon in standard electrodynamics. The right-handed side terms arise from the extra dimension introduced earlier. However, the first term provides the usual fifth-dimensional component, which comes solely from the existence of the extra dimension. The second one provides a correction term due to a small perturbation in the photon energy caused by the Lorentz invariance violation (LIV) term. It is possible for $\alpha_A^2\ll1$, which is the interpretation considered and expected.

\begin{figure}[!htb]
\includegraphics[scale=0.25]{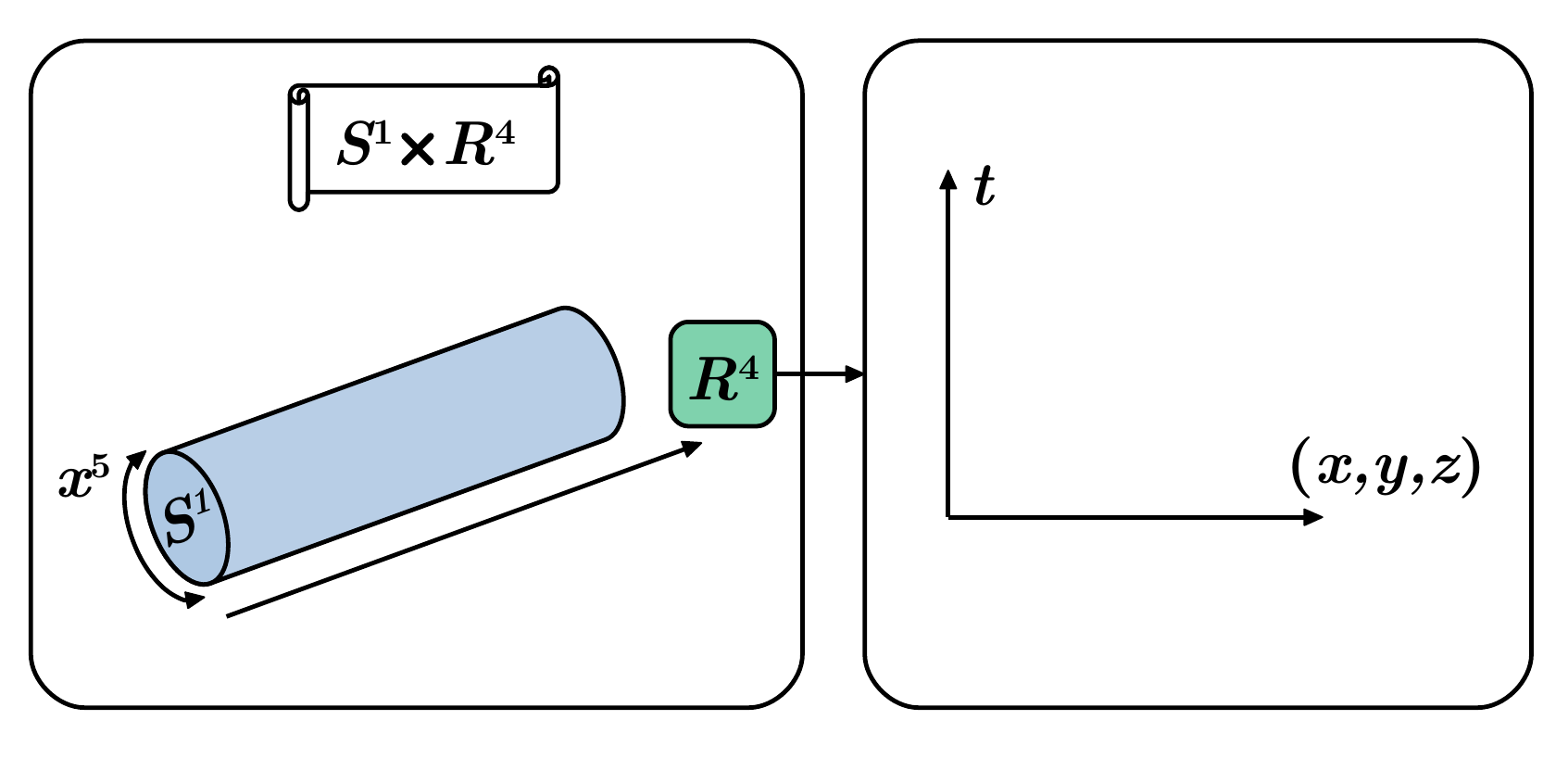}
\caption{Illustrative view of the compactified extra dimension $x^5$.}
\label{fig001}
\end{figure}

If we now impose the quasi-periodic boundary condition
\begin{equation}
A^{a}(x^5)=e^{-2 \pi \beta i} A^{a}(x^5+b),
    \label{eq05}
\end{equation}
as well as the Neumann boundary condition
\begin{equation}
\partial_zA_{a}|_{z=0}=0,\left.\quad \partial_z A_{a}\right|_{z=a}=0,
    \label{eq06}
\end{equation}
on the plane wave solution of the gauge field, the momenta in the $x_5$ and $z$ directions will be discretized.  Note that the quasiperiodic boundary condition is characterized by the phase introduced in terms of one parameter which is limited in the interval $0 \leq \beta < 1$. It is a generalization of the well-known periodic and antiperiodic boundary conditions represented by $\beta=0$ and $\beta=1/2$, respectively. The extra dimension, thus, is compactified into a circle $\mathbb{S}^1$ of length $b$, by the quasiperiodic condition. An illustrative view of this is shown in Fig.\ref{fig001}

\begin{figure}[!htb]
\includegraphics[scale=0.2]{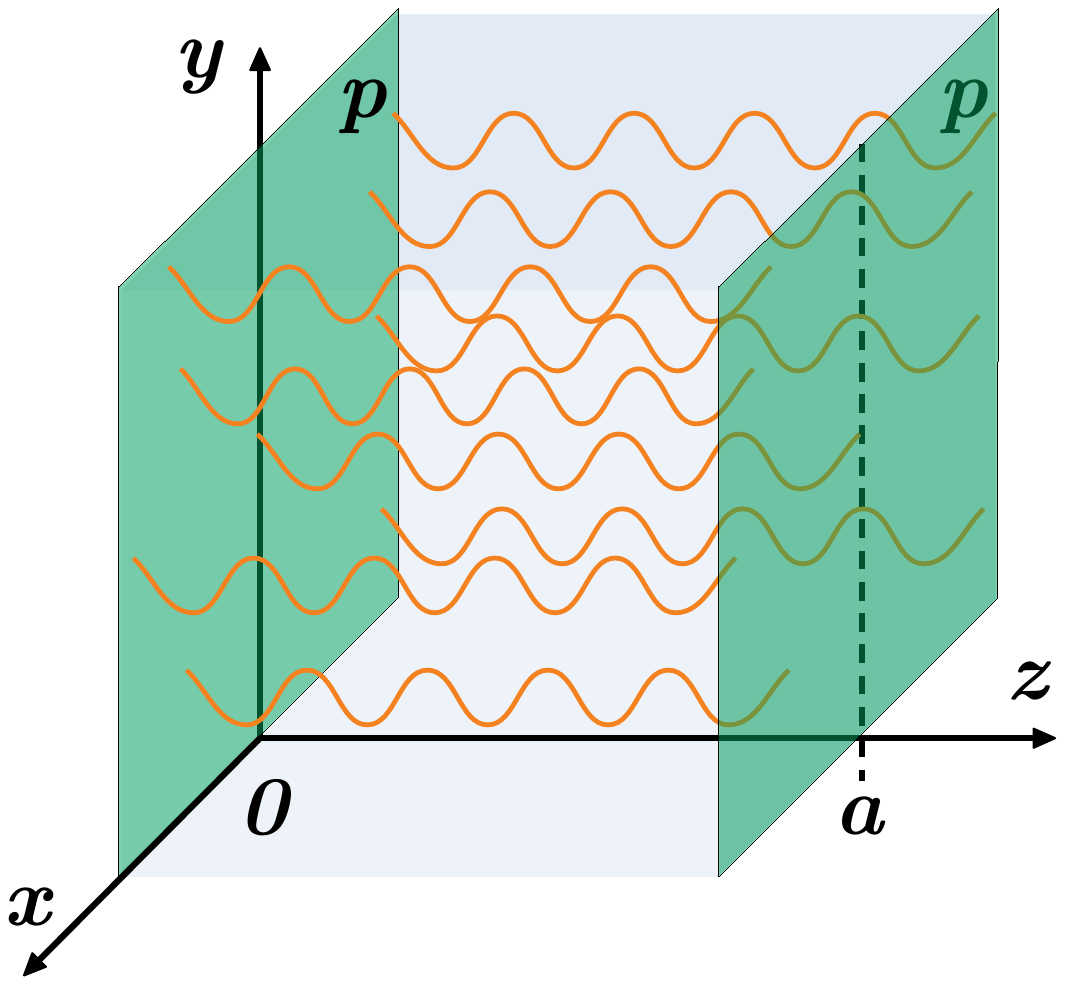}
\caption{Illustrative view of the Neumann boundary condition applied on two parallel plates.}
\label{fig002}
\end{figure}

The presence of the Neumann boundary condition in the $ z$-direction, on the other hand, becomes important since the Casimir effect can be measured in laboratory experiments. As a consequence, the effects of the extra-dimension may be detected as modifications in the Casimir energy. We have also presented an illustration of this in Fig.\ref{fig002}.

 Here we are working in the $\mathbb{R}^4\times\mathbb{S}^1$ spacetime, where $\mathbb{R}^4$ represents the Minkowski spacetime, and $\mathbb{S}^1$ represents the compactified dimension, in our case, the fifth dimension. Therefore, its associated dispersion relation submitted to both boundary conditions can be written as
\begin{equation}
    -k^\mu k_\mu=k_m^2+\left(1+\alpha_A^2\right)k_n^2,
        \label{eq07}
\end{equation}
where $k_m=\frac{m \pi}{a}$ with $m=0,1,2,\ldots$ is related to the Neumann boundary condition where $a$ is the distance between the plates, and $k_n=\frac{2 \pi}{b}(n+\beta)$ with $n=(0,\pm1,\pm2,\ldots)$ is related to the quasi-periodic boundary condition, being $b$ the length of the compactification in $z$-direction. We should point out that the wave momentum discretization  $k_m=\frac{m \pi}{a}$ has the same form as the Dirichlet boundary condition, but in that case $m=1,2,\ldots$, it occurs due to the difference in the modes of the field $\sin$ and $\cos$ function to Dirichlet and Neuman boundary conditions respectively. However, both give us the same physical energy \cite{ambjorn1983properties}. Consequently, the eigenfrequencies are written as
\begin{equation}
\omega_n^2=k_x^2+k_y^2+k_m^2+\left(1+\alpha_A^2\right)k_n^2.
\label{EF1}
\end{equation}
The expression above shall be used to calculate the Casimir energy in the next section.

\section{Casimir energy}
\label{sec3}

As we have said before, in flat spacetime, the Casimir energy arises as a consequence of boundary conditions imposed on the quantum field. As we have imposed on the system two distinct boundary conditions \eqref{eq07}, a finite Casimir energy contribution can be evaluated. In this way, the total vacuum energy is given by the following expression \cite{Poppenhaeger:2003es,Bordag:2009zz}
\begin{equation}
  E=\frac{1}{2}\left(\frac{L}{2 \pi}\right)^{\ell} \int d^{\ell} k \sum_{n=-\infty}^{\infty} \left[p\sum_{m=0}^{\infty} \sqrt{k^2+k_m^2+\left(1+\alpha_A^2\right)k_n^2}+(-p+1)\sqrt{k^2+\left(1+\alpha_A^2\right)k_n^2}\right],
  \label{e1}
\end{equation}
where in our case $k^2=k_x^2+k_y^2$, $L^2$ is the area of the plates and $\ell=2$. Here, $p$ is the possible polarizations of the photon (In the case of extra dimension $p=3$) and for $n = 0$, there is only one polarization possible \cite{Bordag:2009zz}. The parameter $\ell$ is called the dimension regularization. We use it instead of the dimension number because it allows us to perform the integral \eqref{e1} through the relation
\begin{equation}
    \int f(k) d^n k=\frac{2 \pi^{n / 2}}{\Gamma(n / 2)} \int k^{n-1} f(k) d k.
\end{equation}
Consequently, the integral in \eqref{e1} may be solved by writing it as
\begin{align}
E=&\frac{1}{2}\left(\frac{L}{2 \pi}\right)^{\ell} \frac{2 \pi^{\ell / 2}}{\Gamma(\ell / 2)}  \sum_n \int d k k^{\ell-1}\left[ p\sum_m[k^2+k_m^2+\left(1+\alpha_A^2\right)k_n^2]^{-r} +(-p+1)[k^2+\left(1+\alpha_A^2\right)k_n^2]^{-r}\right]\nonumber\\
=&\left(\frac{L}{2 \pi}\right)^{\ell} \frac{\pi^{{\ell} / 2} \Gamma(r-\ell/2)}{2 \Gamma(r)} \sum_n\left\{p\sum_{m}\left[\left(\frac{m \pi}{a}\right)^2+\left(1+\alpha_A^2\right)\left[\frac{2 \pi}{b}(n+\beta)\right]^2\right]^{\frac{\ell}{2}-r}\nonumber\right.\\
&\left.+(-p+1)\left[\left(1+\alpha_A^2\right)\left[\frac{2 \pi}{b}(n+\beta)\right]^2\right]^{\frac{\ell}{2}-r}\right\},
\label{e1.1}
\end{align}
where $r=-1/2$ is a regularization parameter that has allowed us to compute the integral above, it is removed by taking the limit $r\to-1/2$. Now we  replace the exponent on the brackets by $s=r-\ell/2$, and we get
\begin{equation}
    E =\left(\frac{2}{L}\right)^{2s+1} \frac{\pi^{s+1/ 2} \Gamma(s)}{2 \Gamma(-1/2)}\left(\frac{\pi}{a}\right)^{-2 s}\sum_{n=-\infty}^{\infty}\left\{ p\sum_{m=0}^{\infty} \left[\frac{a^2}{\Tilde{b}^2}(n+\beta)^2+m^2\right]^{-s}+(-p+1)\left[\frac{a}{\Tilde{b}}(n+\beta)\right]^{-2s}\right\},
\label{e2}
\end{equation}
%
with
\begin{equation}
\Tilde{b}^2=\frac{b^2}{4\left(1+\alpha_A^2\right)}.
\label{LR}
\end{equation}
As argued in Ref.  \cite{Carroll:2008pk} the aether parameter $\alpha_A$ can be as large as $10^{15}$ and, in this sense, it acts as a length reduction according to Eq. \eqref{LR}.

Note that in Eq \eqref{e2} there are two sums. We will choose the sum in $m$ in order to apply the Abel-Plana formula \cite{Saharian:2007ph} given by the following expression:

\begin{equation}
\sum_{n=0}^{\infty} f(n)=\int_0^{\infty} d x f(x)+\frac{1}{2} f(0)+i \int_0^{\infty} d x \frac{f(x)-f(-i x)}{e^{2 \pi x}-1}.
\label{e3}
\end{equation}
This method shows the divergent contribution that comes from the Minkowski spacetime and also enables us to obtain a finite contribution to the Casimir energy. By first considering  the substitution
\begin{equation}
    f(m)=\left[\frac{a^2}{\Tilde{b}^2}(n+\beta)^2+m^2\right]^{-s}=\left[N^2+m^2\right]^{-s},
    \label{e4}
\end{equation}
from \eqref{e2} and \eqref{e3}, we obtain,
\begin{align}
\sum_{m=0}^{\infty} f(m) =\int_0^{\infty} d x\left(N^2+x^2\right)^{-s}+\frac{N^{-2 s}}{2}+2 i^{1-2 s} \int_N^{\infty} d x \frac{\left(x^2-N^2\right)^{-s}}{e^{2 x \pi}-1}.
\label{e5}
\end{align}
Now, upon using the relation $\left(e^{2 x}-1\right)^{-1}=\sum_{n=1}^{\infty} e^{-a n x}$ in the second integral above, we have
\begin{equation}
    \int_N^{\infty} d x \frac{\left(x^2-N^2\right)^{-s}}{e^{2 x \pi}-1}=\sum_{q=1}^{\infty} \int_N^{\infty} d x e^{-2 x \pi q}\left(x^2-N^2\right)^{-s}=\sum_{q=1}^{\infty}\left(\frac{q}{N}\right)^{-\frac{1}{2}+s} \pi^{-1+s} K_{-\frac{1}{2}+s}  \left(2 N q \pi\right) \Gamma(1-s),
    \label{e6}
\end{equation}
where $K_{\alpha}(x)$ is the modified Bessel function of the second kind. Finally, by substituting \eqref{e6} in Eq. \eqref{e5}, the Casimir energy takes the form
\begin{align}
E= & \left(\frac{2}{L}\right)^{2s+1} \frac{\pi^{s+1/ 2} \Gamma(s)}{2 \Gamma(-1/2)}\left(\frac{\pi}{a}\right)^{-2 s} \sum_{n=-\infty}^{\infty}\left\{p\int_0^{\infty} d x\left(N^2+x^2\right)^{-s}+(-p+2)\frac{N^{-2 s}}{2}+\right.\nonumber \\
& \left.+2 i^{1-2s} \pi^{-1+s} \Gamma(1-s)p \sum_{q=1}^{\infty}\left(\frac{q}{N}\right)^{-\frac{1}{2}+s} K_{-\frac{1}{2}+s}(2 N q \pi)\right\} .
\label{e7}
\end{align}
Note that the above expression includes a divergent contribution that comes from the integral term. However, let us first perform the sum in $n$ to extract a finite contribution, also present in the integral above. After that, we can proceed to discard the divergent contribution by means of the renormalization process. Thus, for the integral term, we have
\begin{equation}
    \int_0^{\infty} d x \sum_{n=-\infty}^{\infty}\left(N^2+x^2\right)^{-s}=\int_0^{\infty} d x\left(\frac{a}{\Tilde{b}}\right)^{-2s} \sum_{n=\infty}^{\infty}\left[(n+\beta)^2+\frac{\Tilde{b}^2}{a^2} x^2\right]^{-s}.
    \label{e8.0}
\end{equation}
The sum may be computed by using the expression \eqref{a6}, providing

\begin{align}
\left(\frac{a}{\Tilde{b}}\right)^{-2s}\!\!\!\! \int_0^{\infty}\!\!\!\! d x\!\!\!\! \sum_{n=-\infty}^{\infty}\left[(n+\beta)^2+\frac{\Tilde{b}^2}{a^2} x^2\right]^{-s}\!\!\!\!
 =&\left(\frac{a}{\tilde{b}}\right)^{-2 s} \int_0^{\infty} d x\left[2\tilde{b}^{-2 s}\int_0^{\infty}d\mu \left(\frac{\mu^2}{\tilde{b}^2}+\frac{x^2}{a^2}\right)^{-s}\right. \\ \nonumber\\
& \left.+\sin(\pi s)
\frac{4\Gamma(1-s)}{\pi^{1-s}}\left(\frac{\Tilde{b}}{a} x\right)^{\frac{1}{2}-s} \sum_{n=1}^{\infty} \frac{\cos (2 \pi n \beta)}{n^{\frac{1}{2}-s}} K_{\frac{1}{2}-s}\left(2 \pi n x\frac{\Tilde{b}}{a}\right)\right].
\label{e8}
\end{align}
Hence, the integral in $\mu$ gives a divergent result, and should be removed \cite{ambjorn1983properties, Bordag:2009zz}. The integral in $x$, on the other hand, can be evaluated by using the following identity:
\begin{equation}
    \int_0^{\infty} d x\left(\frac{b}{a} x\right)^{\frac{1}{2}-s} K_{\frac{1}{2}-s}\left(2 \pi n \frac{b}{a} x\right)=\frac{1}{4}\left(\frac{a}{b}\right) n^{-\frac{3}{2}-s} \pi^{-1+s} \Gamma\left(1-s\right).
    \label{e10}
\end{equation}

Consequently, the result of Eq. \eqref{e8.0} is given by
\begin{equation}
\left(\frac{a}{\Tilde{b}}\right)^{-2s} \int_0^{\infty} d x \sum_{n=-\infty}^{\infty}\left[(n+\beta)^2+\frac{\Tilde{b}^2}{a^2} x^2\right]^{-s}=\left(\frac{a}{\Tilde{b}}\right)^{1-2s} \frac{\pi^{-1+2s}}{\Gamma(s)} \Gamma\left(1-s\right) \sum_{n=1}^{\infty} \frac{\cos (2 \pi n \beta)}{n^{2-2s}},
 \label{e11}
 \end{equation}
where we have used the Euler's reflection formula
\begin{equation}
\Gamma(1-z) \Gamma(z)=\frac{\pi}{\sin \pi z}.
\end{equation}

The second sum of Eq. \eqref{e7} gives an analytical expression, and it is obtained by following the recurrence relation $(A.4)$ from \cite{Feng:2013zza}:
\begin{align}
\sum_{n=-\infty}^{\infty}\left(\frac{a}{\Tilde{b}}\right)^{-2 s}(n+\beta)^{-2 s} =\left(\frac{a}{\Tilde{b}}\right)^{-2s} \pi^{2s-1/2} \frac{\Gamma\left(\frac{1}{2}-s\right)}{\Gamma\left(s\right)} 2 \sum_{n=1}^{\infty}\frac{\cos (2 \pi n \beta)}{ n^{1-2s}} .
\label{e12}
\end{align}

By substituting \eqref{e11} and \eqref{e12} in Eq. \eqref{e7}, we finally arrive at the closed form of the regularized Casimir energy which is given by
\begin{align}
E^{reg}(a,\beta)=&-\left(\frac{2}{L}\right)^{2s+1} \frac{\pi^{s} \Gamma(s)}{4}\left(\frac{\pi}{a}\right)^{-2s} \left\{\left(\frac{a}{\Tilde{b}}\right)^{1-2s} \frac{\pi^{-1+2s}}{\Gamma(s)} \Gamma\left(1-s\right) p\sum_{n=1}^{\infty} \frac{\cos (2 \pi n \beta)}{n^{2-2s}}\right.\nonumber\\
&+(-p+2)\left(\frac{a}{\Tilde{b}}\right)^{-2s} \pi^{2s-1/2} \frac{\Gamma\left(\frac{1}{2}-s\right)}{\Gamma\left(s\right)} 2 \sum_{n=1}^{\infty} n^{2s-1} \cos (2 \pi n \beta)\nonumber\\
&\left.+2 i^{1-2s} \pi^{-1+s} \Gamma(1-s) p\sum_{n=-\infty}^{\infty}\sum_{q=1}^{\infty}\left(\frac{q}{N}\right)^{\frac{-1}{2}+s} K_{\frac{-1}{2}+s}(2 N q \pi)\right\}.
\label{e13}
\end{align}

By considering $s=-3/2$, we remove the regularization parameter imposed on the energy and obtain the following result for the renormalized Casimir energy:
\begin{align}
E^{ren}(a,\beta)=&-\frac{L^2\pi^2}{4a^3}\left\{\frac{3}{16}\left(\frac{a}{\pi \Tilde{b}}\right)^4 p\sum_{n=1}^{\infty} \frac{\cos (2 \pi n \beta)}{n^5}+\frac{-p+2}{2\pi^{4}}\left(\frac{a}{\Tilde{b}}\right)^3 \sum_{n=1}^{\infty} \frac{\cos (2 \pi n \beta)}{n^4}\right.\nonumber\\
&\left.+\frac{p}{2 \pi^2} \sum_{n=-\infty}^{\infty} \sum_{q=1}^{\infty}\left(\frac{N}{q}\right)^2 K_2(2 N q \pi)\right\}.
\label{e14}
\end{align}
It is worth highlighting that, due to the quasi-periodic parameter $\beta$, we do not need to remove the $n=0$ term in the last sum. This term will be analyzed on another occasion when we consider the limit $\beta\to0$. The influence of the quasi-periodic parameter $\beta$ can be found in each term in the brackets. Note that the second term in brackets is not influenced by $a$. If $\beta=0$ is taken, the influence of the quasiperiodic boundary condition vanishes, but this term remains, which means that this term exists only due to the presence of the extra dimension.

Alternatively, we can perform the sum in $n$ present in the first two terms of Eq. \eqref{e14}. This gives
\begin{align}
E^{ren}(a,\beta)= & -\frac{L^2}{8 a\Tilde{b}^2} \sum_{\delta=+,{-}}\left\{p\frac{3}{16}\left(\frac{a}{\pi \Tilde{b}}\right)^2 Li_5\left(e^{2 i \pi \delta \beta}\right)+ (-p+2)\frac{a}{2\pi^{2} \Tilde{b}} Li_{4}\left(e^{2 i \pi \delta \beta}\right)\right.\nonumber\\
&\left.+p\sum_{n=1}^{\infty} \sum_{q=1}^{\infty}\left(\frac{\delta n+\beta}{q}\right)^2 K_2\left[\frac{a}{\Tilde{b}}|\delta n+\beta| 2 q \pi\right]\right\}-p\frac{ L^2}{8a\Tilde{b}^2} \sum_{q=1}^{\infty}\left(\frac{\beta}{q}\right)^2 K_2\left[2\beta q \pi\frac{a}{\Tilde{b}}\right],
\label{e15}
\end{align}
where the summation in $\delta$ means that the parameter $\delta$ assumes two values, $\delta=1,-1$.
The function $Li_s(z)$ is the polylogarithm function, and the last term of the above expression corresponds to the $n=0$ term.

Note that, in Eq. \eqref{e15}, the first term in brackets is linear in $a$. In Ref. \cite{pascoal2008estimate}, in a similar study, the authors disregard this term because when the limit $a\to\infty$ is taken, it diverges. However, this term contributes as a constant in the force, the real quantity to be measured. Since our aim is to investigate the influence of the boundary condition and the extra-dimension parameter in the Casimir force, we will consider this term, and analyze how it affects the force. Thus, this term cannot be ignored in the experiment. 

\begin{figure}[!htb]
\includegraphics[scale=0.295]{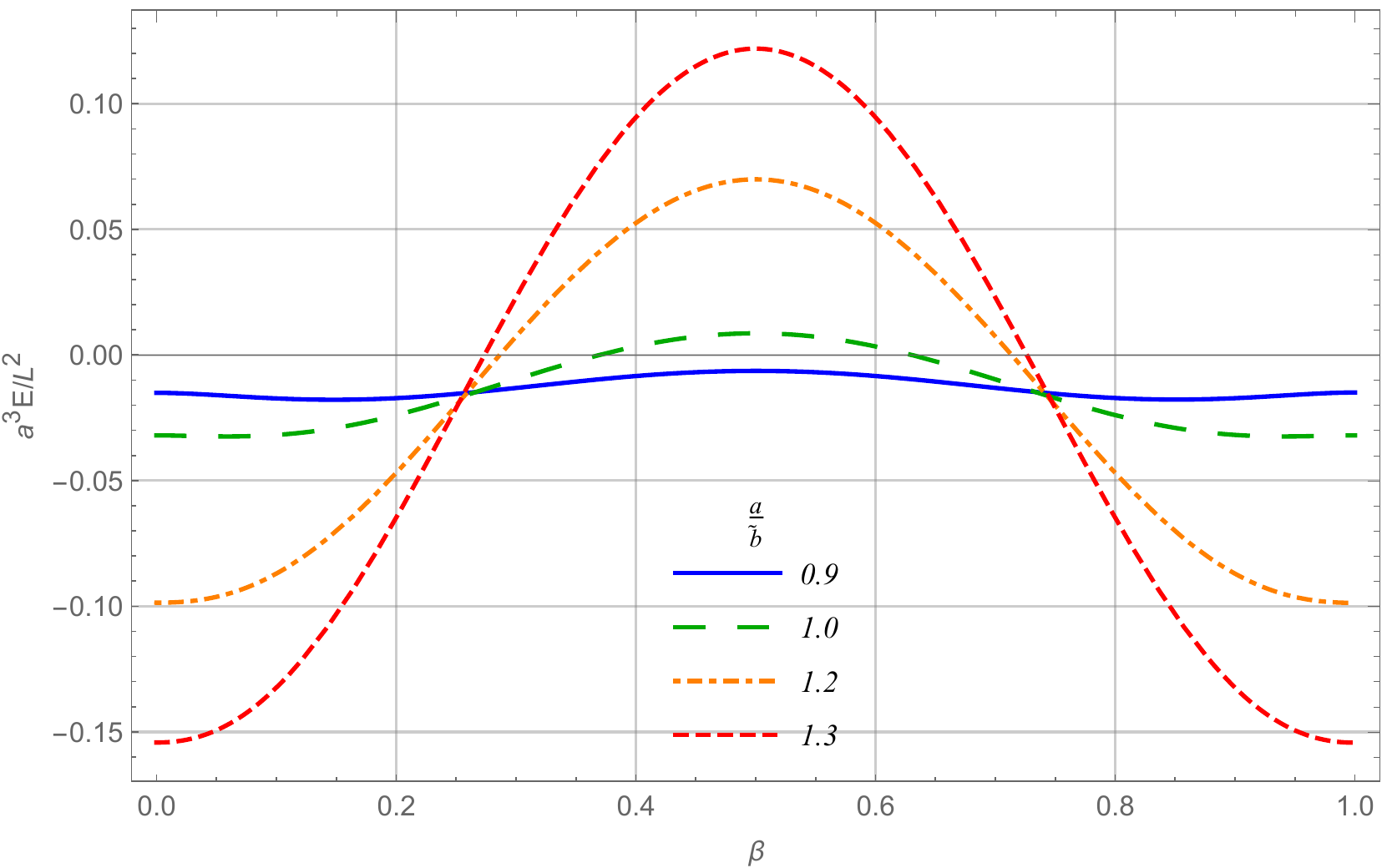}
\includegraphics[scale=0.3]{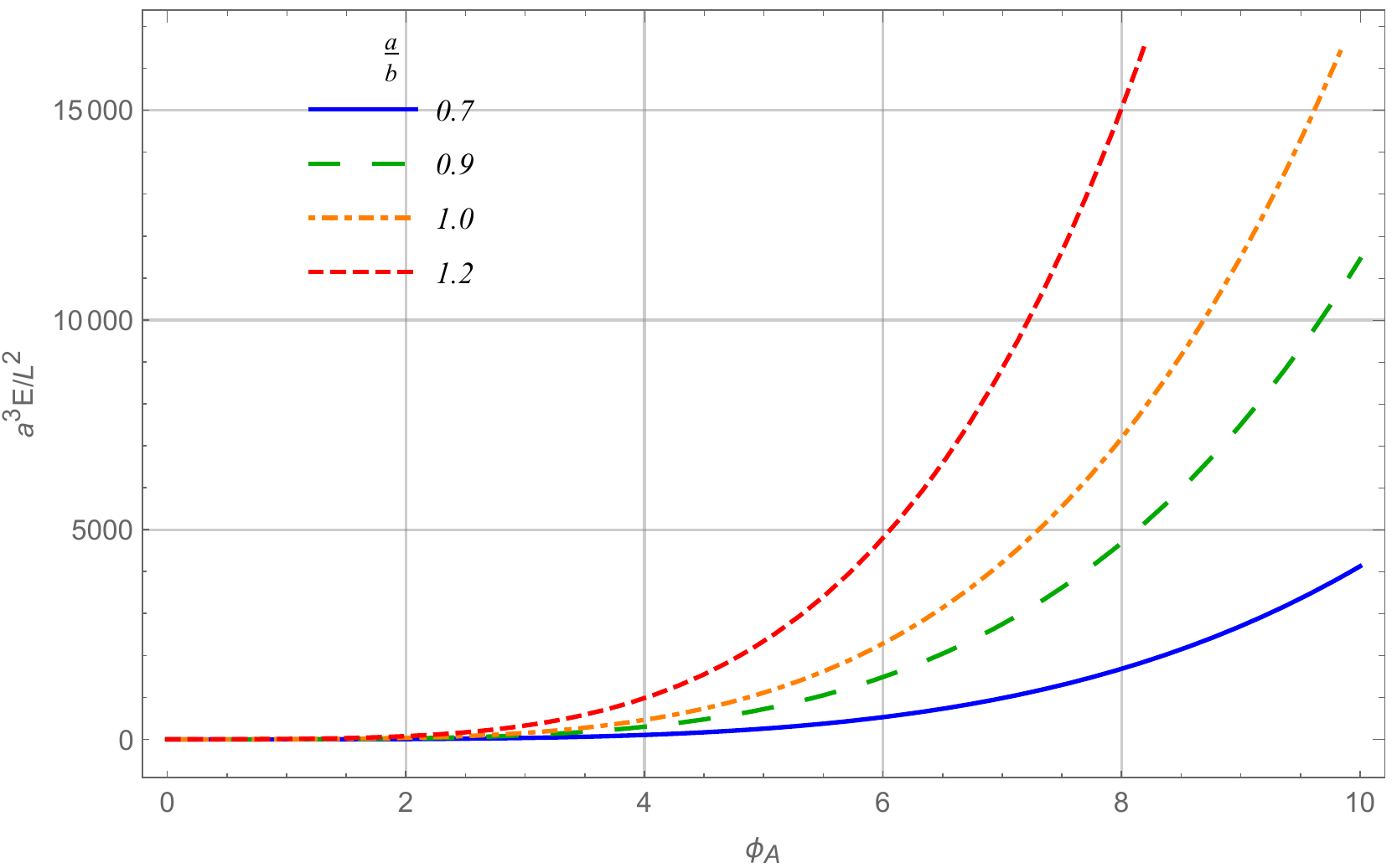}
\caption{The energy density $a^3E/L^2$, for several values of the ratio $a/\Tilde{b}$, in terms of $\beta$ (left) and $a/b$, in terms of $\alpha_{A}$ (right). In the right we consider $\beta =0.4$.}
\label{fige02}
\end{figure}

\begin{figure}[!htb]
\includegraphics[scale=0.3]{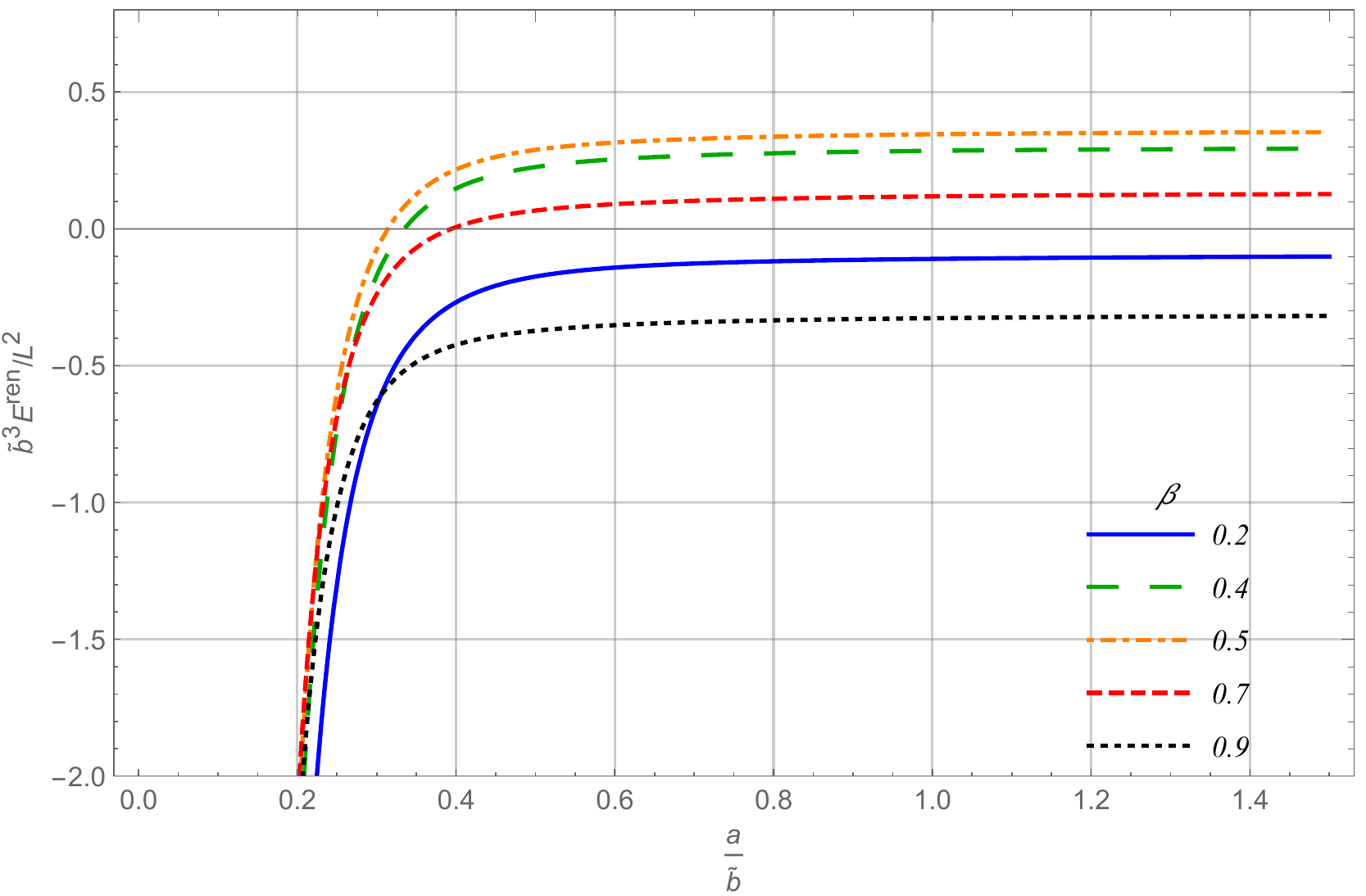}
\caption{The energy density $\Tilde{b}^3E^{\rm ren}/L^2$, for several values of the ratio $\beta$, in terms of the ratio $a/\Tilde{b}$. There is a different notation, $E^{\rm ren}$, in this figure, when compared to the previous due to the first term of the r.h.s. of \eqref{e15} which diverges as $a\to\infty$. Naturally, it should be removed to get the correct energy density behaviour}
\label{fig003}
\end{figure}

In Fig.\ref{fige02}, we have plotted the Casimir energy \eqref{e15} per unit of area of the plates, with respect to the parameters $\beta$ and  $\alpha_{A}$. On the left, the plot shows that $\beta$ may be used to control the intensity and sign of the vacuum energy. This means that we can have an attractive, repulsive or even zero vacuum energy. Moreover, we can see in the plot that the value $\beta=1/2$ presents a symmetry, giving each curve the maximum positive value for the vacuum energy. The values $\beta=0,1$, also present a symmetry, providing the same negative minimum value for the vacuum energy for each curve. We can see that the intensity of the vacuum energy increases as $a/\Tilde{b}$ also increases. Note that, considering a numerical analysis, the energy becomes positive in the approximated interval $0.25< \beta< 0.75$, regardless of the values of $a/\Tilde{b}$. On the other hand, the influence of the parameter $\alpha_{A}$ is shown in the right of Fig.\ref{fige02}, which reveals that the energy per unit area of the plates is divergent as $\alpha_{A}\to\infty$. Hence, a nonzero aether parameter should increase the renormalized Casimir energy density. However, since there exists in principle an upper limit for the parameter $\alpha_{A}$ \cite{Carroll:2008pk}, the vacuum energy will never be divergent in practice. In Fig.\ref{fig003} we have also plotted the Casimir energy \eqref{e15} per unit of area of the plates, with respect to the ratio $a/\Tilde{b}$. It is clear that when this ratio goes to zero the energy diverges while when it goes to infinity the energy goes to a constant value for each curve.
%

Now, we will consider the case where $\beta=0$, in which the quasiperiodic boundary condition is absent, and we recover the periodic condition. Indeed, the last term of \eqref{e15} represented by $n=0$ is divergent if we directly substitute $\beta=0$. However, we can take the limit $\beta\to0$ for the quasiperiodic parameter, which results in a finite contribution as follows
\begin{equation}
\sum_{q=1}^{\infty}\left(\frac{\beta}{q}\right)^2 K_2\left(2 \beta q \pi\frac{a}{\Tilde{b}}\right) \stackrel{\beta \rightarrow 0}{\longrightarrow} \sum_{q=1}^{\infty} \frac{\Tilde{b}^2}{2a^2 \pi^2 q^4}=\frac{\pi^2\Tilde{b}^2}{180a^2}.
\label{e15.1}
\end{equation}

By making use of the above result in Eq. \eqref{e15}, we find
\begin{equation}
E^{ren}(a,0)=-\frac{L^2 \pi^2}{1440 a^3}\left[p+(-p+2)\frac{2a^3}{ \Tilde{b}^3}+p\frac{135 a^4}{2 \pi^4 \Tilde{b}^4} \zeta(5)+p\frac{360 a^2}{\pi^2 \Tilde{b}^2}\sum_{n=1}^{\infty} \sum_{q=1}^{\infty} \frac{n^2}{q^2} K_2\left(2 n \pi q \frac{a}{\Tilde{b}}\right)\right],
\label{e15.2}
\end{equation}
which is in accordance with Refs. \cite{pascoal2008estimate, Zhai:2014jta}, showing the consistency of the results obtained so far. We can note that the first term in the r.h.s. of the above expression is the Casimir energy due to a gauge field subjected to a Neumann boundary condition in (3+1)-dimensional Minkowski spacetime ($p=2$) \cite{ambjorn1983properties}, while the second term is a correction due to the extra dimension only since it does not depend on the parameter $a$. However, this term will not contribute to the force between the plates, as we shall see in the next section. The third and fourth terms are corrections due to both, the presence of the plates and the compactified extra dimension.

\section{Casimir force}
\label{sec4}

 Although we have obtained an analytical expression for the vacuum energy, in Eq. \eqref{e15}, our focus is on finding the force acting on the plates in the $z-$direction. The importance of this quantity stays from the possibility of analyzing the correction terms in Eq. \eqref{e15} by making use of experimental data. Let us then calculate the Casimir force first. This is given by
\begin{align}
 F_a(a, \Tilde{b})=&-\frac{\partial E(a, \Tilde{b})}{\partial a}=-p\frac{L^2}{128 \Tilde{b}^4} \sum_{\delta=+,-}\left\{-\frac{3}{\pi^2} Li_5\left(e^{2 i \pi\beta\delta}\right)\right.\nonumber \\
& +16\sum_{n=1}^{\infty} \sum_{q=1}^{\infty}\left[3\left(\frac{ \Tilde{b}(\delta n+\beta)}{aq}\right)^2 K_2\left(\frac{a}{\Tilde{b}} 2 \pi q|\delta n+\beta|\right)+2\pi\left.\frac{ \Tilde{b} (\delta n+\beta)^3}{a q} K_1\left(\frac{a}{\Tilde{b}} 2 \pi q|\delta n+\beta|\right)\right]\right\}\nonumber \\
&-p\frac{L^2}{8 \Tilde{b}^4}\sum_{q=1}^{\infty}\left[3\left(\frac{ \Tilde{b}\beta}{aq}\right)^2 K_2\left(\frac{a}{\Tilde{b}} 2 \pi q\beta\right)+2\pi\frac{ \Tilde{b} \beta^3}{a q} K_1\left(\frac{a}{\Tilde{b}} 2 \pi q\beta\right)\right].
\label{e16}
\end{align}
This force arises directly from the Casimir energy in Eq. \eqref{e15}. Hence, we may associate this quantity with a Casimir force that acts due to the insertion of the boundary conditions in the system. Note that in the r.h.s, the first term in the above expression refers to the pure contribution of the compactified extra dimension, since there is no influence of $a$. Note also that the last term is due to the contribution $n=0$ from the sum in $n$.

\begin{figure}[!htb]
\includegraphics[scale=0.3]{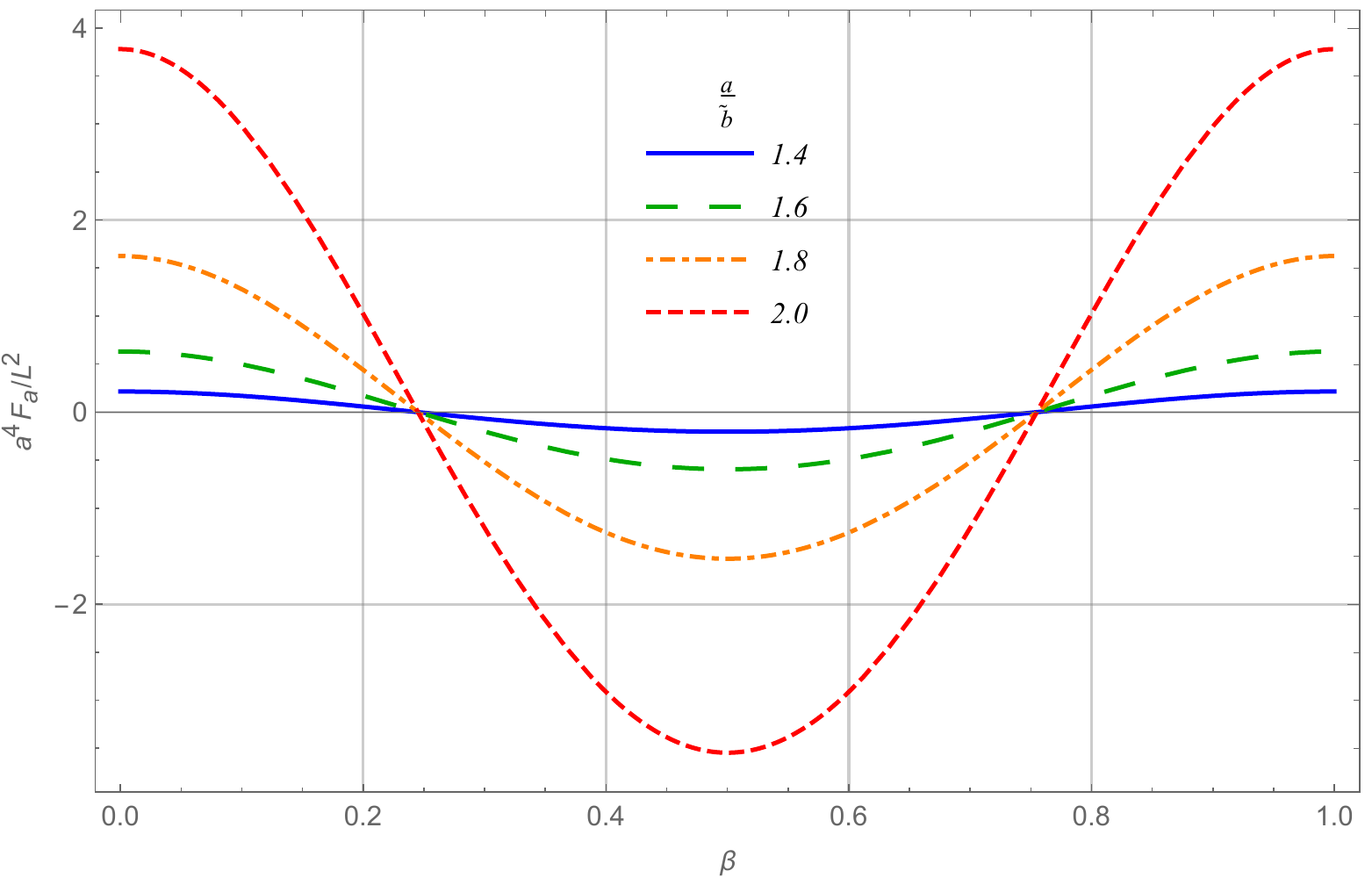}
\includegraphics[scale=0.3]{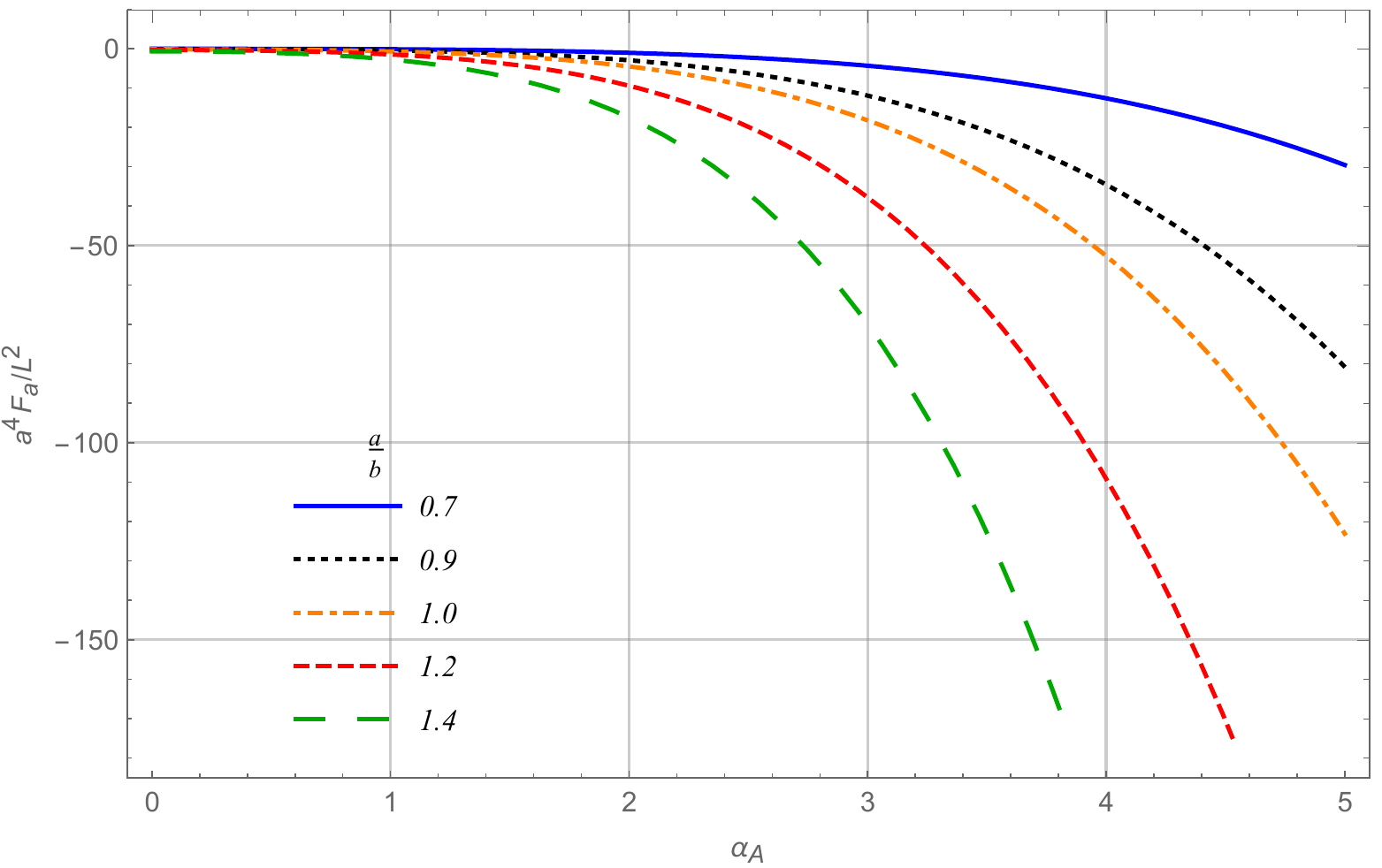}
\caption{The force $a^4F_a/L^2$, for several values of the ratio $a/\Tilde{b}$, in terms of $\beta$ (left) and $a/b$ in terms of $\alpha_{A}$ (right). In the right we consider $\beta =0.4$.}
\label{fige0}
\end{figure}

In Fig.\ref{fige0} we have plotted the Casimir force in Eq. \eqref{e16}, per unit area of the plates, in terms of the parameters $\beta$ (left) and $\alpha_{A}$ (right). We can see that  $\beta$ has a clear influence on the nature (attractive, repulsive or zero) of the force, as well as on its intensity. Similar behaviour can be found in the investigation conducted, for instance, in Refs. \cite{deFarias:2020xms,deFarias:2021qdg}, reinforcing that the parameter $\beta$ can be used to control the nature of the force. We can also see that the force diverges as $\alpha_{A}$ becomes larger.

\begin{figure}[!htb]
\includegraphics[scale=0.3]{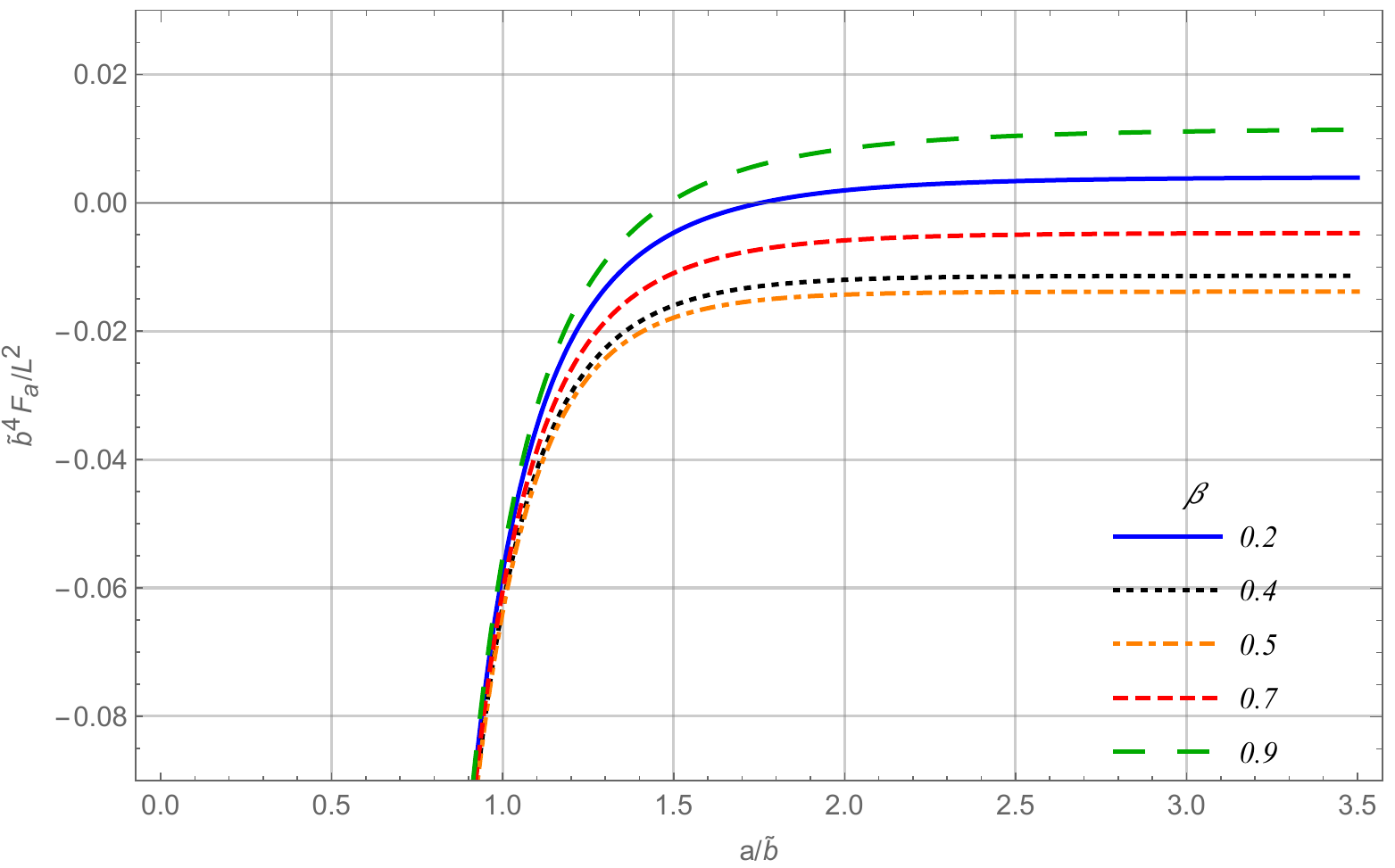}
\caption{The force $a^4F_a/L^2$, for several values of the ratio $\beta$, in terms of the ratio $a/\Tilde{b}$.}
\label{fige01}
\end{figure}

In Fig.\ref{fige01}, on the other hand, it is shown the behaviour of the Casimir force, per unit area of the plates, in terms of the ratio $a/\Tilde{b}$. The plot shows that for small values of this ratio, the Casimir force negatively increases (up to infinity), while for large values the Casimir force goes to a constant positive value for each curve shown.

\subsection{Specific values}

Let us now consider the important cases of the periodic $(\beta=0)$ and antiperiodic $(\beta=1/2)$ conditions. By starting with the former case, the Casimir force \eqref{e16} is written as
\begin{equation}
    F_a(a,\Tilde{b})=-p\frac{L^2 \pi^2}{480a^4}+p\frac{3L^2 }{64 \Tilde{b}^4\pi^2} \zeta(5)- p\frac{L^2\pi}{2a\Tilde{b}^3}\sum_{n=1}^{\infty} \sum_{q=1}^{\infty}\frac{ n^3}{q} K_1\left(2\pi n q \frac{a}{\Tilde{b}}\right)- p\frac{3L^2}{(2a\Tilde{b})^2}\sum_{n=1}^{\infty} \sum_{q=1}^{\infty}\left(\frac{ n}{q} \right)^2 K_2\left(2\pi n q \frac{a}{\Tilde{b}}\right),
   \label{e20}
\end{equation}
where we have made use of Eq. \eqref{e15.1}. The above result agrees with the expression obtained in Ref. \cite{pascoal2008estimate}. One should note that, by taking the limit $a\to\infty$, we obtain the extra dimension contribution to the Casimir force, that is,
 \begin{equation}
      F_a(\infty,\Tilde{b})=p\frac{12\zeta(5)L^2 }{\pi^2b^4}(1+\alpha_{A})^2 ,
 \end{equation}
which is in agreement with Ref. \cite{ambjorn1983properties} in the absence of the aether parameter, i.e., when $\alpha_{A}=0$. In the limit $a\to0$, on the other hand, the force \eqref{e15.1} diverges, as expected.


Following the special considerations, another interesting value to investigate separately is $\beta=1/2$. This specific value leads us to the antiperiodic condition case. From Eq.\eqref{e20}, we obtain:
\begin{align}
F_a(a, \Tilde{b})= & -p\frac{45 L^2}{1024 \Tilde{b}^4} \zeta(5)-p\frac{L^2}{32 a \Tilde{b}^3} \sum_{\delta=+,-}\sum_{n=1}^{\infty} \sum_{q=1}^{\infty}\left\{\frac{\pi}{q}(2 \delta n+1)^3 K_1\left[\frac{a}{\Tilde{b}} \pi q|2 \delta n+1|\right]\right.\nonumber \\
& \left.+\frac{\Tilde{b}(2 \delta n+1)^2}{3 a q^2} K_2\left[x \pi q(2 \delta n+1)\right]\right\} -p\frac{L^2}{32 a \Tilde{b}^3} \sum_{q=1}^{\infty}\left[\frac{\pi}{q} K_1\left(\frac{a \pi q}{\Tilde{b}}\right)+\frac{3 \Tilde{b}}{a q^2} K_2\left(\frac{a \pi q}{\Tilde{b}}\right)\right].
\label{e21}
\end{align}
In this case, the force diverges negatively as $a\to0$ and, as it is shown in Fig.\ref{fige01}, remains always attractive. Hence, the presence of an extra-dimension parameter $\Tilde{b}$ affects the force, causing an attractive force between the plates.


\section{Constraint to the size of the extra-dimension}
\label{sec5}

There are currently many models that attempt to impose constraints and describe the properties and size of extra dimensions \cite{Murata:2014nra}. However, these models pose significant challenges for experimental verification due to the sensitivity required to detect extra dimensions of small size. As mentioned earlier in this work, according to the Kaluza-Klein theory, the size of the extra dimension would be close to the Planck scale, making it hard, not to say impossible, to make direct measurements using the current experimental apparatus. Here, however, we propose an indirect method to detect the existence of this extra dimension and derive accessible constraints on both the size of the extra dimension and the two new parameters introduced in our model: the aether ratio $\alpha_{A}$ and the quasi-periodicity parameter $\beta$.

Indeed, the experiments related to the Casimir effect remain very difficult, even with the development of technology. The first attempt to prove the Casimir effect was made by Sparnaay in 1958 \cite{Sparnaay:1958wg}, but the measure errors found in the experiments could not confirm the effect. Later, in 1997, Lamoreux proved the existence of the Casimir effect \cite{Lamoreaux:1996wh}, which shows this force experimentally with acceptable accuracy. However, there is a point that we should highlight. The force measured by the experiments does not identify the origin of the force. A good point of view of the Casimir force comes from the Jaffe's interpretation of this force. In his paper \cite{Jaffe:2005vp}, he shows that the Casimir energy does not come from the zero point energy (modifications in the quantum vacuum), but he computes the Casimir force by using relativistic, quantum forces between charges and currents. Therefore, he concludes there is no need to use the zero-point energy or vacuum fluctuations argument when treating Casimir force.

Based on Jaffes's work, the Casimir effect is a function of the fine structure constant $\alpha$, and it vanishes as $\alpha\to0$, as well as any observable effect in QED. In the experiment developed by Bressi et al., \cite{Bressi:2002fr}, they compare the experimental data obtained with the result found by Casimir \cite{Casimir:1948dh} for the parallel conducting plates
\begin{equation}
F=-\frac{\hbar c \pi^2}{240 a^4},
\end{equation}
where $a$ is the distance between the plates. According to Jaffe \cite{Jaffe:2005vp}, the above result is an asymptotic form in the limit $\alpha \rightarrow \infty$, where the use of boundary condition forces this limit. Consequently, the explicit dependence of $\alpha$ cannot be seen, which leads us to a semiclassical limit. As a consequence of this independence, the Casimir force \eqref{e16} also represents a semiclassical result associated with vacuum fluctuations of quantized fields. For simplicity, we consider that the  Casimir effect is reduced to a modification of the spectrum of standing waves of a classical field between ideally conducting plates, in a way to compare with the experimental data in the semiclassical approach, where $\alpha\to\infty$. Hence, the contribution of the fine structure constant exists, but it does not influence the result. Although the Casimir effect cannot be reduced to a modification of the spectrum of standing waves of a classical field, our aim in this section is not only to investigate the influence of the boundary condition on manipulating the result through the variable parameters but also show the possibility to estimate a constraint to the size of the extra dimension.

Following the method proposed in Ref. \cite{pascoal2008estimate}, we can estimate the size of the extra dimension by taking the derivative of the pressure with respect to the distance between the plates, denoted as $a$, as shown in Ref. \cite{Bressi:2000iw}. That is,
\begin{equation}
    \Delta \nu^2=\nu^2-\nu_0^2=-\frac{L^2}{4 \pi^2 m_{\text{eff}}} \frac{\partial P}{\partial a}=-\frac{L^2}{4 \pi^2 m_{\text{eff}}}  \frac{\partial}{\partial a}\left[\frac{F_a(a, \tilde{b})}{L^2}\right].
    \label{e22}
\end{equation}
where $m_{\text{eff}}$ is an effective mass related to the properties of the system and $L^2$ represents the effective area between the cantilever and the source surfaces where the fluctuation occurs. Measuring the effective mass is not straightforward, but as suggested in Ref. \cite{Bressi:2000iw}, the ratio $L^2/m_{\text{eff}}$ can be estimated using the experimental data. This is approximately given by $L^2/m_{\text{eff}} \approx 1.746 \text{~Hz}^2 \text{m}^3 \text{N}^{-1}$ \cite{Bressi:2000iw}.

Eq. \eqref{e22} leads to the square plates' oscillating frequency shift, denoted as $\Delta \nu^2$. This is the quantity that the experimental apparatus can detect. For more details about the experiment, see Refs. \cite{Bressi:2000iw} and \cite{Bressi:2002fr}. Subsequently, by substituting the result \eqref{e16} into Eq. \eqref{e22}, we obtain:
\begin{align}
\Delta \nu^2   =&-p\frac{\hslash cL^2}{4 \pi^2 m_{\text{eff}}} \sum_{n=-\infty}^{\infty} \sum_{q=1}^{\infty} \frac{(n+\beta)^2}{2 a^3 \tilde{b}^2 q^3}\left\{\left[3+\pi^2 q^2(n+\beta)^2 \frac{a^2}{\tilde{b}^2}\right] K_0\left(2 \pi q|n+\beta| \frac{a}{\tilde{b}}\right)\right.\nonumber \\
& \left.+\frac{\tilde{b}}{2 \pi q a|n+\beta|}\left[6+5 \pi^2 q^2(n+\beta)^2 \frac{a^2}{\tilde{b}^2}\right] K_1\left(2 \pi q|n+\beta| \frac{a}{\tilde{b}}\right)\right\}
\label{e23}
\end{align}
Note that we have recovered the constants $(\hslash,c)$ at this point since they are essential for comparing the model with the experimental. Consequently, by making use of Eq. \eqref{e23}, it is possible to estimate the size of the extra dimension. This estimation will also have an impact on the quasiperiodic parameter as well as the aether parameter. This is advantageous because the presence of these parameters may assist in establishing an acceptable limit for the length $b$.

In a scenario where the parameter $\beta$ is equal to $0$, we obtain the following expression:
\begin{align}
\Delta \nu^2   =& -p\frac{\hslash cL^2}{4 \pi^2 m_{\text{eff}}}\left\{\frac{\pi^2}{120 a^5}+\sum_{n=1}^{\infty} \sum_{q=1}^{\infty} \frac{1}{a^3 b^2}\left[\left(\frac{3 n^2}{q^2}+\frac{a^2 \pi^2 n^4}{\tilde{b}^2}\right) K_0\left(\frac{2a \pi q n}{\tilde{b}}\right)+\right.\right.\nonumber \\
& \left.\left.+\frac{\tilde{b}}{2 \pi}\left(\frac{3 n}{q^3}+\frac{5 \pi^2}{2} \frac{a^2}{\tilde{b}^2} \frac{n^3}{q}\right) K_1\left(\frac{2a \pi q n}{\tilde{b}}\right)\right]\right\}.
\label{e24}
\end{align}
The above expression matches the result obtained in Ref. \cite{pascoal2008estimate}, where it is acknowledged that this result is not suitable for determining the size of the extra dimension. However, the model studied in Ref. \cite{pascoal2008estimate} does not consider the quasiperiodic $\beta$ or the aether parameter in the calculation. Therefore, by incorporating the experimental data from Ref. \cite{Bressi:2002fr}, constraints on $\alpha_{A}$ can be imposed to determine a compatible length $b$.

\begin{figure}[h]
    \centering
    \includegraphics[scale=0.4]{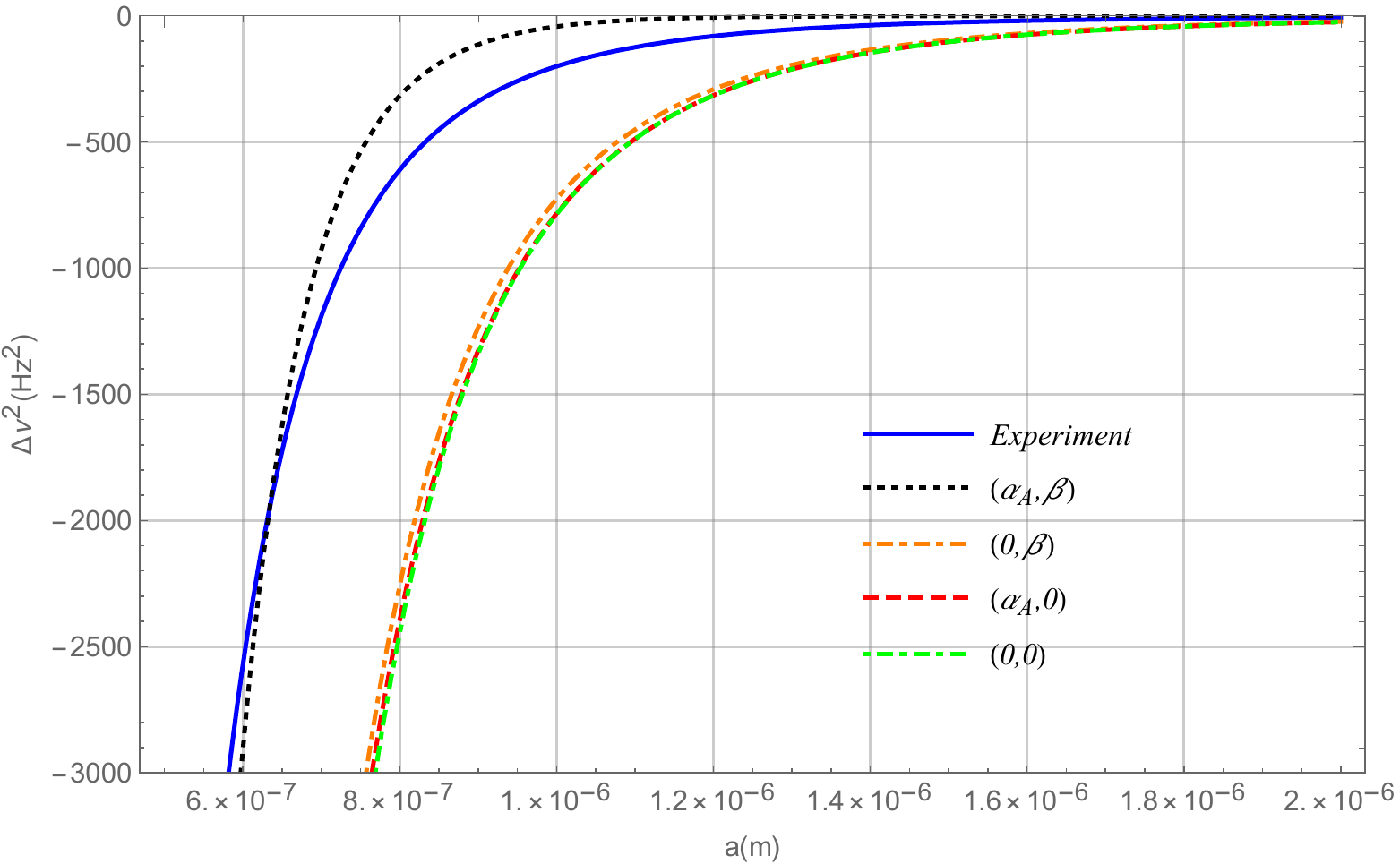}
    \caption{The residual square frequency shift, $\Delta\nu^2$, is shown as a function of the plate distance, $a$, for $b=1.0\times10^{-6}$ $m$, $\alpha_{A}=10^{4}$, and $\beta=0.006$. The solid line represents the experimental data from Ref. \cite{Bressi:2002fr}. The lines where $\beta\neq0$ come from Eq. \eqref{e23}, and the lines with $\beta=0$ represent the behavior of Eq. \eqref{e24}.}
    \label{fenom}
\end{figure}

Following Ref. \cite{Bressi:2002fr}, the residual squared frequency shift obtained by the experiment is given by:
\begin{equation}
\Delta \nu^2(a)=-\frac{C_{\mathrm{Cas}}}{a^5},
\label{e25}
\end{equation}
where $C_{\text{Cas}} = (2.34 \pm 0.34) \times 10^{-28} \text{~Hz}^2 \text{~m}^5$, which is also experimentally determined \cite{Bressi:2000iw}. With all the parameters in hand, we can now incorporate the experimental data from \cite{Bressi:2002fr}. This is shown in Fig.\ref{fenom}.

Fig.\ref{fenom} illustrates the influence of the parameters in comparison to the experimental data. Although the aether ratio parameter affects the frequency shift when $\beta=0$, it does not show any significant enhancement compared to the case with $\alpha_{A}=0$. It's evident that the best result is achieved when both $\alpha_{A}$ and $\beta$ are not equal to zero, yielding an approximate size for the extra dimension, $b=1.0\times10^{-6}\ m$, with $\alpha_{A}=10^4$ and $\beta=0.006$. It is clear that this value for $b$ aligns with the range of the cantilever experiment \cite{Murata:2014nra}.

Let us now point out an important detail related to the role of $\alpha_{A}$. As mentioned below Eq. \eqref{LR}, the aether parameter acts as a length reduction which can make the detection of the size of the extra dimension become more challenging. However, it still remains crucial to align the theoretical model with experimental data, as shown in Fig.\ref{fenom}. The theoretical fit presented in Fig.\ref{fenom} indicates that it is not a perfect match within a certain interval. This discrepancy can be attributed to external factors not considered in our theoretical model, including finite temperature corrections, finite conductivity of the material used in the experiment, and other factors detailed in \cite{Bressi:2002fr}.

The results obtained in this paper represent the first instance of such findings and serve to extend the results previously documented in the literature. Additionally, new constraints have been identified for the size of the extra dimension within a LIV aether scenario in the context of the Casimir effect.


\section{Conclusion}\label{sec6}

In this work, we have investigated the influence of the aether compactification as a means of explaining the size problem of the fifth dimension in flat spacetime, in conjunction with the quasiperiodic parameter $\beta$. The potential existence of an extra dimension modifies the dispersion relation of the standard theory. This extra dimension has been compactified into length $b$, and we have also introduced two parallel plates in the $z$-direction in order to apply the Neumann boundary condition. In this sense, both the compactified extra dimension and the Neumann boundary condition applied on the plates lead to the Casimir effect.

The modified Casimir energy density has been found in a closed form, expressing the energy in terms of the quasiperiodic parameter $\beta$, the distance between the plates $a$, the length of the extra dimension $b$, and the aether parameter $\alpha_{A}$. We have shown several plots to illustrate the behaviour of the energy under the influence of the quasiperiodic parameter $\beta$, which are shown in Fig.\ref{fige02}, \ref{fig003}. These plots reveal the sign of the energy and the values for $\beta$ where it vanishes.

The direct proportionality of the energy in terms of the aether parameter has been demonstrated in Fig. \ref{fig003}, showing an increase in energy as $\alpha_{A}\to\infty$. We have taken the limit $\beta\to0$, where the influence of the quasiperiodic parameter vanishes. In this limit, the result obtained is consistent with the literature \cite{pascoal2008estimate, Zhai:2014jta}, and the well-known result for Casimir energy from the Neumann boundary condition emerges \cite{ambjorn1983properties}. Additionally, pure terms from the extra dimension and mixed terms among the parameters become apparent. The energy, when influenced by the quasiperiodic parameter, oscillates between negative and positive values depending on the value of $\beta$, while for $\beta=0$, the energy remains entirely negative.

We have found a closed expression for the force acting on the plates in the $z$-direction in terms of $\beta$, $a$, and $\Tilde{b}$ (Eq. \eqref{e16}). We have also plotted the force as a function of $\beta$ for various values of $a/\Tilde{b}$ (Fig.\ref{fige0}), in terms of $\alpha_{A}$ (Fig.\ref{fige0}) and in terms of $a/\Tilde{b}$ (Fig.\ref{fige01}). The intensity of the force is controlled by the choice of $a/\Tilde{b}$, $\beta$ and $\alpha_{A}$, as shown in the plots. Similar to the energy, $\alpha_{A}$ contributes to the growth of the force between the plates (Fig.\ref{fige0}). We have also seen that the force approaches a constant value as $a/\Tilde{b}\to\infty$ (Fig.\ref{fige01}).

We have observed that the energy diverges as $\alpha_{A}\to\infty$ in Fig.\ref{fige02}. However, the aether parameter is limited, so the energy has an upper limit in terms of $\alpha_{A}$. Furthermore, in our investigation of the Casimir force, we have considered the limits of the force for $\beta\to0$, which aligns with the literature \cite{pascoal2008estimate}, and for $\beta=1/2$, which corresponds to the antiperiodic boundary condition. In the case of the periodic boundary condition, as $\beta\to0$, we recover the standard result for the Casimir force influenced by the Neumann boundary condition, as found in \cite{ambjorn1983properties}.

Finally, we have tested our model using experimental data obtained in \cite{Bressi:2002fr} to place an upper limit on the size of the extra dimension $b$ and the aether parameter $\alpha_{A}$. The model demonstrates consistency in recovering the result from the reference work \cite{pascoal2008estimate} when $\beta\to0$, as represented by Eq. \eqref{e24}. From Eq. \eqref{e23}, we have created Fig.\ref{fenom}, which provides information about the behaviour of the system and indicates that the only viable result arises from the values $\alpha_{A},\beta\neq0$, yielding a size of $b=1.0\times10^{-6}m$, and for the aether parameter, we find a value of $\alpha_{A}=10^4$ when considering $\beta=0.006$. The constraint for the size falls within the range of the cantilever experiment, confirming its ability to provide evidence of the existence of a Lorentz invariance violation (LIV) scenario and an extra dimension.

In conclusion, the results for the Casimir energy \eqref{e15}, the Casimir force \eqref{e16}, and the square frequency shift \eqref{e23} in a scenario with quasiperiodic boundary condition, together with an LIV parameter, have been developed for the first time in this work. This generalizes investigations available in the literature.


{\acknowledgments}

We would like to thank CNPq and CAPES for partial financial support. K.E.L.F would like to thank the Brazilian agency CNPq for financial support. M.A.A, and E.P acknowledge support from CNPq (Grant nos. 306398/2021-4, 304852/2017-1, Universal 406875/2023-5). J.R.L.S would like to thank CNPq (Grant nos. 420479/2018-0, and 309494/2021-4), and PRONEX/CNPq/FAPESQPB (Grant nos. 165/2018, and 0015/2019) for financial support. H.F.S.M is partially supported by CNPq under grant no. 311031/2020-0.


\appendix

\section{Sum of $\sum f(n+\beta)$}

Now, we deduce the recurrence formula to solve the integral \eqref{e8.0}. Starting with the Abel-Plana formula as follows
\begin{equation}
    \sum_{n=0}^{\infty} f(n+a)=\int_0^{\infty} f(x) d x+\frac{f(a)}{2}-i \int_0^{\infty}dx \frac{f(a + i x)-f(a-i x)}{e^{2 i x}-1}
    \label{a0}
\end{equation}

The sum can be divided into positive and negative intervals:
\begin{equation}
\sum_{n=-\infty}^{\infty} f(n+\beta)=\sum_{n=0}^{\infty} f(n+\beta)+\sum_{n=0}^{-\infty} f(n+\beta)-f(\beta)=\sum_{n=0}^{\infty} f(n+\beta)+\sum_{n=0}^{\infty} f(-n+\beta)-f(\beta)
\label{a1}
\end{equation}
Applying the sums separately in the Abel-Plana formula \eqref{a0}, we have
\begin{equation}
     \sum_{n=0}^{\infty}\left[(n+s)^2+\mu^2\right]^{-s}=\int_0^{\infty} d x\left[x^2+\mu^2\right]^{-s}+\frac{\left(\beta^2+\mu^2\right)^{-s}}{2} -i \int_0^{\infty} dx\frac{\left[(\beta+i x)^2+\mu^2\right]^{-s}-\left[(\beta-i x)^2+\mu^2\right]^{-s}}{e^{2 \pi x}-1}
     \label{a2}
\end{equation}
 and

\begin{align}
\sum_{n=0}^{\infty}\left[(-n+\beta)^2+\mu^2\right]^{-s}= & \sum_{n=0}^{\infty}\left[(n+(-\beta))^2+\mu^2\right]^{-s}\nonumber \\
= & \int_0^{\infty} d x\left[x^2+\mu^2\right]^{-s}+\frac{\left[(-\beta)^2+\mu^2\right]^{-s}}{2}\nonumber\\
&-i \int_0^{\infty} dx\frac{\left.\left.[((-\beta)+i x)^2+\mu^2\right]^{-s}-[((-\beta)-i x)^2+\mu^2\right]^{-s}}{e^{2 \pi x}-1}\nonumber \\
= & \int_0^{\infty}\!\!\!\! d x\left[x^2+\mu^2\right]^{-s}+\frac{\left(\beta^2+\mu^2\right)^{-s}}{2}-i \int_0^{\infty}\!\!\!\!dx \frac{\left[(-\beta+i x)^2+\mu^2\right]^{-s}-\left[(-\beta-i x)^2+\mu^2\right]^{-s}}{e^{2 \pi x}-1}
\label{a3}
\end{align}

By substituting \eqref{a2} and \eqref{a3} in Eq. \eqref{a1}, we get
\begin{align}
\sum_{n=-\infty}^{\infty} \left[(n+\beta)^2+\mu^2\right]^{-s}=&\int_0^{\infty} d x\left[x^2+\mu^2\right]^{-s}+\frac{\left(\beta^2+\mu^2\right)^{-n}}{2}-i \int_0^{\infty} d x \frac{\left[(\beta+i x)^2+\mu^2\right]^{-s}-\left[(\beta-i x)^2+\mu^2\right]^{-s}}{e^{2 n x} - 1} \nonumber\\
& +\int_0^{\infty} d x\left[x^2+\mu^2\right]^{-s}+\frac{\left(\beta^2+\mu^2\right)^{-s}}{2}\nonumber\\
& -i \int_0^{\infty} d x \frac{\left[(-\beta+i x)^2+\mu^2\right]^{-s}-\left[(-\beta-i x)^2+\mu^2\right]^{-s}}{e^{2 n x} - 1} -\left(\beta^2+\mu^2\right)^{-s} \nonumber\\
 =&2 \int_0^{\infty} d x\left[x^2+\mu^2\right]^{-s}+ \left(i^{1-2s}+(-i)^{1-2s}\right)\int_0^{\infty} d x \frac{\left[(x+i \beta)^2-\mu^2\right]^{-s}+\left[(x-i \beta)^2-\mu^2\right]^{-s}}{e^{2 \pi x}-1}
\label{a4}
\end{align}

The second integral of \eqref{a4} may be solved by using the identity
\begin{equation}
\left(e^y-1\right)^{-1}=\sum_{j=1}^{\infty} e^{-j y} .
\end{equation}

As a consequence, we have

\begin{equation}
\begin{aligned}
\sum_{n=-\infty}^{\infty}\left[(n+\beta)^2+\mu^2\right]^{-s}=&2 \int_0^{\infty} d x\left[x^2+\mu^2\right]^{-s}\nonumber\\
&+\left(i^{1-2s}+(-i)^{1-2s}\right) \sum_{\ell=1}^{\infty} \int_0^{\infty} d x\left(\left[(x+i \beta)^2-\mu^2\right]^{-s}+\left[(x-i \beta)^2-\mu^2\right]^{-s}\right) e^{-2 \ell\pi x} \\
& =2 \int_0^{\infty}\!\!\!\! d x\left[x^2+\mu^2\right]^{-s}\nonumber\\
&+\left(i^{1-2s}+(-i)^{1-2s}\right) \sum_{\ell=1}^{\infty}\left\{\int_0^{\infty}\!\!\!\! d u e^{-2 \pi \ell(u-i \beta)}\left[u^2-\mu^2\right]^{-s} +\int_0^{\infty}\!\!\!\! d \nu e^{-2 \pi \ell(\nu+i \beta)}\left[\nu^2-\mu^2\right]^{-s}\right\}
\label{a5}
\end{aligned}
\end{equation}

The above expression can be reduced to
\begin{align}
\sum_{n=-\infty}^{\infty}\left[(n+\beta)^2+\mu^2\right]^{-s}&=2 \int_0^{\infty} d x\left[x^2+\mu^2\right]^{-s}+\left(i^{1-2s}+(-i)^{1-2s}\right) \sum_{\ell=1}^{\infty} 2 \cos (2 \pi \ell \beta) \int_\mu^{\infty} d u e^{-2 \pi \ell u}\left[u^2-\mu^2\right]^{-s}\nonumber \\
&= \sqrt{\pi} \mu^{1-2 s} \frac{\Gamma\left(-\frac{1}{2}+s\right)}{\Gamma(s)}+\left(i^{1-2s}+(-i)^{1-2s}\right)
\frac{2\Gamma(1-s)}{\pi^{1-s}} \sum_{\ell=1}^{\infty}\left(\frac{\ell}{\mu}\right)^{-\frac{1}{2}+s} \cos (2 \pi \ell \beta) K_{\frac{1}{2}-s}(2 \pi \mu \ell)\nonumber\\
&= \sqrt{\pi} \mu^{1-2 s} \frac{\Gamma\left(-\frac{1}{2}+s\right)}{\Gamma(s)}+\sin(\pi s)
\frac{4\Gamma(1-s)}{\pi^{1-s}} \sum_{\ell=1}^{\infty}\left(\frac{\ell}{\mu}\right)^{-\frac{1}{2}+s} \cos (2 \pi \ell \beta) K_{\frac{1}{2}-s}(2 \pi \mu \ell)
\label{a6}
\end{align}
%

\end{document}